\documentclass[10pt,aps,pre,twocolumn,superscriptaddress]{revtex4-2} 
\usepackage[english]{babel}
\usepackage[T1]{fontenc}
\usepackage{tikz}
\usepackage{amssymb,amsmath,amsfonts} 
\usepackage{textcomp} 
\usepackage{braket} 
\usepackage{nicematrix} 
\usepackage{graphicx,color}
\usepackage[dvipsnames]{xcolor}
\usepackage[utf8x]{inputenc}
\usepackage{bm}
\usepackage{mathbbol}
\usepackage{comment}

\usepackage{mathtools}

\usepackage[usenames,dvipsnames,x11names]{xcolor}
\usepackage[normalem]{ulem}
\usepackage{graphicx}
\usepackage{listings}
\usepackage{stmaryrd}
\usepackage{amssymb}
\usepackage{amsmath}
\usepackage{gensymb}
\usepackage{bbold}
\usepackage{bm}

\usepackage[linkcolor = blue, citecolor = red, urlcolor = blue, colorlinks = true]{hyperref}

\usepackage{tabularx}

\begin{document}

%\title{A Clifford algebra description of liquid crystalline fermions} 
%quasiparticles
%Clifford algebras of liquid crystalline fermions
\title{%Active Knotted Polymers as Soft Chiral Particles
Topological localisation and motility of active knots}
%\title{Dynamic Chirality, Reptation and Localisation in Active Polymer Knots}

\author{A. Bonato}
\affiliation{SUPA, School of Physics and Astronomy, Edinburgh EH9 3FD, UK}
\author{D. Marenduzzo}
\affiliation{SUPA, School of Physics and Astronomy, Edinburgh EH9 3FD, UK}
\author{E. Orlandini}
\affiliation{Department of Physics and Astronomy, University of Padova and  Sezione INFN, Padova, Via Marzolo 8, I-35131 Padova, Italy}
\author{G. Negro}
\affiliation{SUPA, School of Physics and Astronomy, Edinburgh EH9 3FD, UK}

\begin{abstract}
Nonequilibrium active polymers provide a minimal framework to investigate biopolymers such as DNA and chromatin under the action of molecular motors. Here we study active ring polymers with controlled topology and show that knot type qualitatively determines their nonequilibrium behaviour. We find that activity induces opposite localisation responses in different topological families: torus knots systematically delocalise and inflate, whereas twist knots tighten and remain localised. We trace this divergent behaviour to the distinct symmetry properties of their tangent fields, which control the alignment of active forces along the chain.
We show that topology also governs internal and emergent dynamics. Active torus knots behave as soft chiral self-propelled particles exhibiting persistent motion with a well-defined handedness fixed by their topological chirality. In contrast,  achiral knots show no net handedness. The knot thus acts as a deformable topological quasiparticle whose morphology and propulsion are selected by topology. These results suggest potential routes toward programmable soft chiral particles with controllable morphology and emergent motility modes.  
\end{abstract}

\maketitle
%\textbf{Introduction}
\section{Introduction}
Biopolymers such as DNA and chromatin are kept out of equilibrium by, amongst others, the continual activity of polymerases~\cite{alberts2015essential,sevier2018properties,forte2026}, molecular motors such as condensin~\cite{alipour2012}, and topological enzymes such as topoisomerases~\cite{bates2005,zidovska2013micron}. Notwithstanding the critical role played by nonequilibrium phenomena, we still have a limited understanding of the effect of activity on polymer dynamics, and chromatin or DNA are still more commonly modelled as systems in thermal equilibrium. 

Over the last decade, minimal models of active polymers have been proposed to explore how internal forcing reshapes polymer conformations and affects their dynamics~\cite{bianco2018,foglino2019,winkler2017,zhu2024,sakaue2026}. Activity has been shown to induce a swollen-to-globule transition driven by persistent propulsion~\cite{bianco2018,breoni2025}, alter the mechanical response by softening the polymer via local kink formation~\cite{foglino2019}, and significantly affect the probability of knot formation in linear diblock copolymers~\cite{vatin2025upsurge}.
In ring geometries, activity can create additional spatiotemporal patterns, leading to rotating or coherently moving states that have no counterpart in equilibrium~\cite{miranda2023,Lamura2024}. More generally, active forcing breaks detailed balance~\cite{prost2015active} and can generate effective tensions, enhanced diffusion, and emergent alignment along the polymer backbone, fundamentally altering the physics of passive polymers~\cite{winkler2017}. For example, dense suspensions of active diblock ring polymers exhibit glassy behavior~\cite{smrek2020} that originates from a network of deadlocks~\cite{micheletti2024topology}. Despite these advances, the role of topology in organizing nonequilibrium active polymers remains largely unexplored.

\textcolor{black}{Understanding this interplay is important not only for biological polymers but also for the design of synthetic active matter. A central challenge in that field is to program motion and transport at the microscale through geometry and the spatial organization of active forces, an avenue opened in particular by advances in 3D microprinting of anisotropic and deformable active particles~\cite{wei2026microprinting}. Shape has already been exploited to encode helical propulsion, while toroidal microswimmers have been engineered to transport passive and active cargo~\cite{baker2019microtori}. Active knots suggest a complementary strategy in which dynamical behavior is encoded not by shape alone, but by topology itself.}

In this work, we study how topology constrains nonequilibrium polymer dynamics by focusing on active knotted rings or \emph{active knots}. These may be viewed as soft topological quasi-particles, similarly to the way in which polymers can be coarse-grained into soft colloids~\cite{likos2006}. In equilibrium, knots can perform anomalous diffusion along the polymer in which they are embedded~\cite{metzler2006} and adjust their size in response to external fields, confinement, or tension~\cite{marcone2005,farago2002pulling,caraglio2015stretching,renner2014untying}. Importantly, different knot families possess distinct symmetry properties~\cite{Adams:1994}.
\textcolor{black}{Torus knots lie on the surface of a torus and are labelled by two coprime integers counting their windings around its two generating cycles; twist knots are built by introducing twists into a closed loop and clasping its ends~\cite{Adams:1994}. Importantly, torus knots are chiral and admit globally ordered tangent fields, providing a potential quasiparticle analogue of active chiral rotors~\cite{carenza2019,carenza2020A,caprini2024}.}

\textcolor{black}{This raises natural questions: how does persistent active forcing reorganise stresses in a topologically constrained chain? Can topology alone encode dynamical chirality and control transport in active polymers? Addressing these questions allows us to treat knots as topology-driven active particles with self-selected morphology and motility.}

\textcolor{black}{We show that activity reshapes polymer knots in a topology-dependent manner. Torus knots systematically delocalise as activity increases, expanding until the knotted region spans the entire ring; this is in sharp contrast to passive polymers under tension, where knots tighten~\cite{arai1999tying,bao2003behavior,farago2002pulling,caraglio2015stretching}. Twist and achiral knots instead become more localised, forming compact entangled regions that slide along a weakly perturbed backbone. In both cases, knots behave as mobile quasiparticles reptating along the ring with a timescale separation between the fast motion of beads through the knotted structure and the slower evolution of the knot's global shape.}

\textcolor{black}{Topology also controls emergent motility: delocalised torus knots self-organise into coherently rotating structures that self-propel via helical trajectories, with handedness fixed by topological chirality. Propulsion efficiency grows with crossing number, making knot type a tunable handle on self-propulsion speed.
Finally, this topological organisation is robust to spatial patterning of activity: regular active–passive patterns with sufficiently small period preserve the torus/twist knot dichotomy, while increasing segment length toward a diblock drives knot migration into the passive block, a crossover from topology-controlled localisation to activity-driven segregation.}

\section{Results}
To study the coupling between topological constraints and activity, we simulate active knots of $N_b = 64$ beads of size $\sigma$ for various knot types $\pi$. This chain length was chosen to accommodate moderately complex knots while preventing the crumpling behavior observed in longer active knots~\cite{breoni2025}. We consider the trefoil knot $3_1$ (which is both a twist and a torus knot), the twist knots $4_1$, $5_2$, $7_2$, the torus knots $5_1$ and $7_1$, and finally the knot $6_3$, which is neither a twist nor a torus knot. %\abn{Check that this is right}.
%\abn{Some quick stuff on activity}
The non-equilibrium properties of the knots are studied via molecular dynamics simulations using LAMMPS~\cite{plim1995}, coupled to a custom modification implementing activity. Each bead in the knotted ring has a diameter $\sigma$ and mass $m$, and evolves according to an overdamped Langevin equation with friction $\zeta=1$ and temperature $T=0.5$ in LJ units (see materials and methods for more details). This sets a unit of time $\tau=\sigma \sqrt{m/2K_BT}$. The beads interact via a WCA potential, while two consecutive beads interact through a FENE potential.
Chain stiffness is described via a bending (Kratky-Porod) potential. 
Knots are assigned an orientation, and activity is modelled as a constant force $F_a$ pushing each bead tangentially to the backbone of the polymer.  We vary the active force  $F_a $  in the interval $F_a \in [0,10]$, in units of $2k_BT/\sigma$, which corresponds to Peclet number $\text{Pe}\coloneqq  F_a \sigma/k_B T \in [0,20]$. More details are given in the Methods section.%For all the results presented we fixed the simulation time-step to $dt=10^{-3}\tau$, equilibrated initial configurations for $300\tau$, and run the simulations for $9\times10^5\tau$ before collecting data and compute observables. More details are given in~\cite{refSI}.
\begin{figure}[t!]
\centering
\includegraphics[width=1\columnwidth]{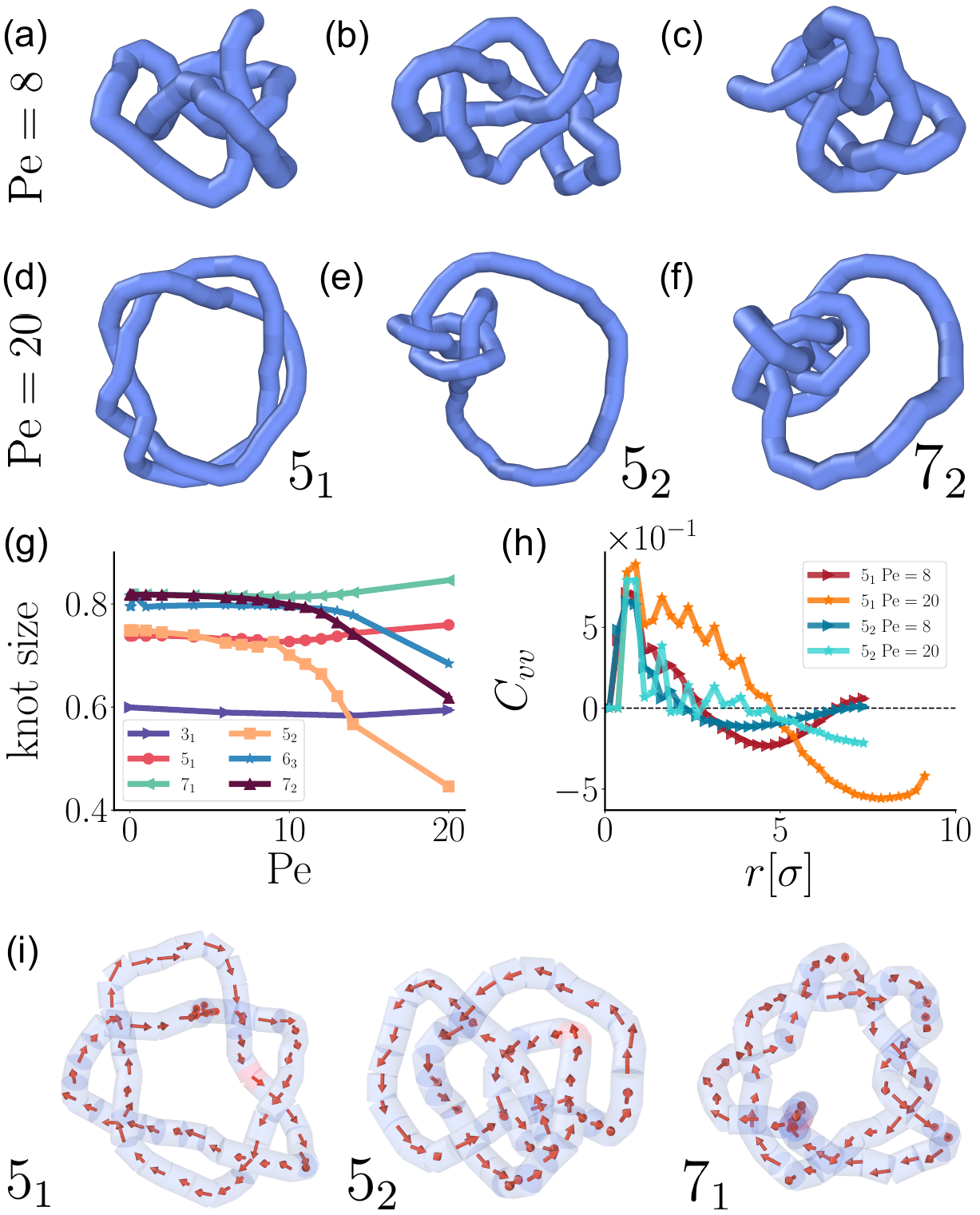}
\caption{\textbf{Localization. Activity can induce a localization or  a delocalization for torus knots.} %Persistence length dell'attivita sull'asola e persistence lenght del nodo rimane quella nominale. Forse diventano piu interessanti le simulazioni al variare della persistence. Vedere il trifoglio perche  e misto, e magari controlli che ala variare della peristencefa una cosa o l'altro 
\textbf{(a-f)} Snapshots from simulations of active knots of different knot types $\pi$, $5_1$ (\textbf{(a),(d)}), $5_2$ (\textbf{(b),(e)}) and $7_2$ (\textbf{(c),(f)}) at two activity strengths, $\text{Pe}=8$ (\textbf{(a),(b),(c)}) and $\text{Pe}=20$ (\textbf{(d),(e),(f)}).
\textbf{(g)} Average fraction of beads in the knotted region as a function of activity for different knot types, averaged at steady state, over different replicas.
\textbf{(h)} Beads velocity-velocity correlation function as a function of \textcolor{black}{distance} in space for $5_1$ and $5_2$ knots simulated at low ($\text{Pe}=8$) and high ($\text{Pe}=20$) activity.
\textbf{(i)} Snapshot of quasi-ideal conformation of a $5_1$, a $5_2$ and a $7_2$ knot from simulations at $\text{Pe}=14$. The knots are decorated with arrows indicating the orientation of the activity. 
%\abn{o facciamo una figura (double column, lineare) a 8/10 pannelli in cu mettiamo pannelli a d b e (c f) h i + nematic correlation, o 2 figure, una a 4/6/8 pannelli in cui mettiamo solo gli snapshots, una a 4 in cui mettiamo h, i + coorelation}
} 
\label{fig1}
\end{figure} 

\subsection{Activity induced knot tightening and delocalisation}
To assess the effects of activity on equilibrated knotted rings, we look at the size and position of the knotted portion. This is done by using \textit{KymoKnot} ~\cite{kymoknot}, which systematically computes the Alexander polynomial~\cite{Adams:1994} on closed subarcs of each sampled configuration~\cite{tubiana2011probing}. 
Remarkably, the behavior of the knotted portion strongly depends on the underlying topology: as the activity increases, the torus knots $5_1$ (Fig.~\ref{fig1}(a)-(d)) and $7_1$ inflate and, at higher values of the active force, adopt quasi-ideal toroidal conformations, where the knotted region occupies the entire ring (Fig.~\ref{fig1}(d) and Movie1)~\cite{katritch1996geometry}. 
Conversely, activity causes the knotted portions of twist knots $5_2$ (Fig.~\ref{fig1}(b)-(e) and Movie2) $7_2$ (Fig.~\ref{fig1}(c)-(f)), and knot $6_3$, to tighten. The resulting conformations can be described as unknotted loops crowned by a sliding root that encapsulates the compact knotted region. Surprisingly, this steady state non-equilibrium structure resembles the conformations adopted by asymptotically long knots and links in the passive regime ~\cite{orlandini1998,bonato2020,bonato2021}.

To quantify the effects of tightening and delocalisation, we computed the average size of the knotted region as a function of the Peclet number. 
As shown in Fig.~\ref{fig1}g, increasing activity causes torus knots to expand beyond their size at equilibrium, whereas other knot types contract significantly. This difference in entanglement localization between torus and non-torus knots can be explained by monitoring the orientation of the knots while diffusing along their contour. When the activity is high, a knot tends to conform to its ideal  shape~\cite{katritch1996geometry}, since it becomes effectively stiffer.
Fig.~\ref{fig1}(i) shows that activity directions are locally disordered in quasi-ideal twist knots but aligned with neighboring beads in torus knots. The polar order emerging in highly active torus knots, which makes them more stable toward big conformational changes, is quantified in Fig.~\ref{fig1}(h), where we report the bead velocity-velocity correlation as a function of distance in space. 
Correlations decay rapidly at both low and high activity for the $5_2$ twist knot, while the $5_1$ torus knot exhibits long-range order at sufficiently high activity. Analogous to nematic or polar ordering and phase transitions in active filament suspensions~\cite{Joshi2019,Athani2024,MirandaLopez2025}, localization and tightening of active knots represent the topological and geometric manifestations of activity-driven organization.
Finally, we consider the trefoil knot $3_1$, which uniquely belongs to both torus and twist knot families. As activity varies, its knot size (Fig.\ref{fig1}(g)) remains delocalized consistently with torus knot behavior. However, the knot size distribution is significantly broader than for pure torus or twist knots (see SI~\cite{refSI}), reflecting its dual topological nature.
%\abn{Secondo me e' piu' chiaro l'argomento della nematicita' che la reptation (che comunque potrebbe avere un ruolo) Direi di spostare lo spiedino in Fig 1 nella figura dopo, concentrarsi su nematic order e suggerire reptation come possibile contributo alla localisation quando spieghiamo la dinamica.}
\begin{figure}[t!]
\centering
\includegraphics[width=1\columnwidth]{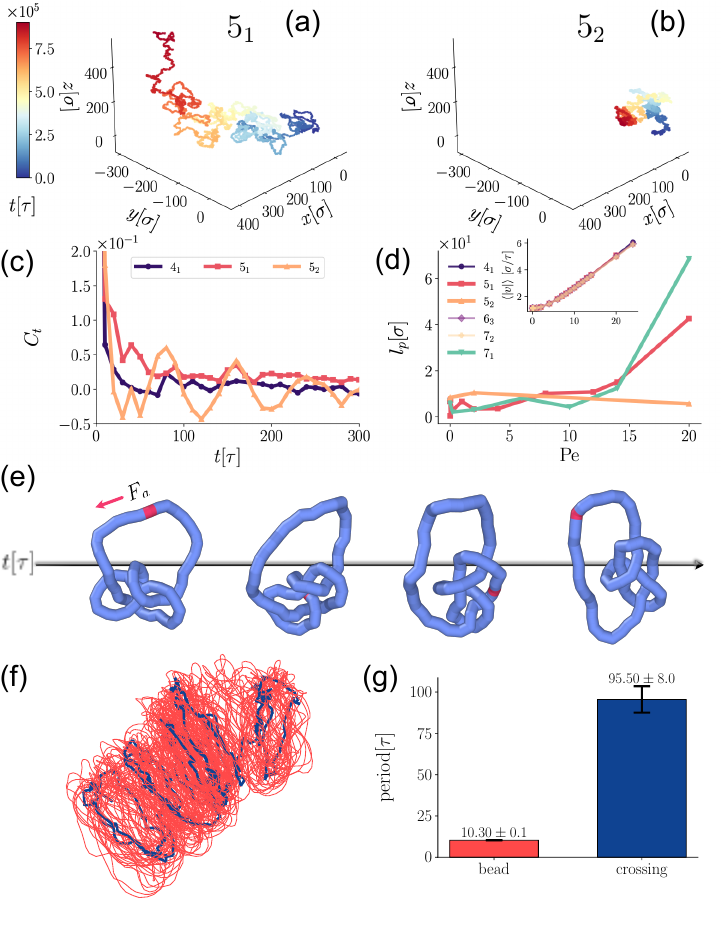}
\caption{\textbf{Persistence and reptation.} 
\textbf{(a-b)} Trajectories of the center of mass of a $5_1$ and a $5_2$ knot simulated at $F_a=10$.
\textbf{(c)} Correlation function (for $\text{Pe}=20$) and persistence length (\textbf{d}) of the tangents to the trajectory of the center of mass of simulated active knots. The inset in panel (\textbf{d}) shows the average velocity of the beads as a function of activity, for different topologies.
\textbf{(e)} Sequence of subsequent snapshots from simulations of a $5_2$ knot revolving in space. Highlighted in red is a given selected bead, which quickly traverses the whole contour before the knot shape changes significantly. (\textbf{f})
Trajectory of the crossing (blue) of a delocalized torus knot ($5_1$ for $\text{Pe}=20$), and of one of its beads (red), over a time interval of $10^3\tau$. \textcolor{black}{\textbf{(g)} Reptation period (red bar) and revolution period of the crossings (blue bar) for the case displayed in panel (f) \textcolor{black}{(see SI for more details on the definition and measurement of these quantities)}}.%Prova integrale per la persistence length.
} 
\label{fig2}
\end{figure}

\subsection{Topology dependent persistent motion of active knots}
The relationship between activity and topology not only manifests through the reported conformational bias but also profoundly affects the dynamics of active knots.
To probe this, we inspected the trajectories of the center of mass of the knot of various knot types for different intensities of activity. 
Figure~\ref{fig2}(a,b) displays representative trajectories for $5_1$ (torus) and $5_2$ (twist) knots at $\text{Pe}=20$ over equal time intervals. 
\textcolor{black}{The torus knot $5_1$ exhibits greater directional persistence than the twist knot $5_2$, resulting in larger spatial displacement, as quantified by the trajectory tangent correlation function (Fig.\ref{fig2}c) and its spatial decay (or persistence) length (Fig.\ref{fig2}d).} Note that the direction of motion is lost over a longer interval of time as the activity increases. 
This effect varies quantitatively with knot type: torus knots exhibit significantly larger motility persistence than other topologies, and within the torus family, higher crossing numbers yield higher persistence. Notably, enhanced persistence does not reflect local bead dynamics—Fig.~\ref{fig2}d inset confirms that average bead velocity is topology-independent. The tangent-tangent correlation function further 
captures some features of the conformations of the knots: in tight knots (e.g., $5_2$ at $\text{Pe}=20$), it oscillates between parallel and antiparallel alignment, whereas it
decays monotonically in all the other cases.

\begin{figure}[t!]
\centering
\includegraphics[width=1.0\columnwidth]{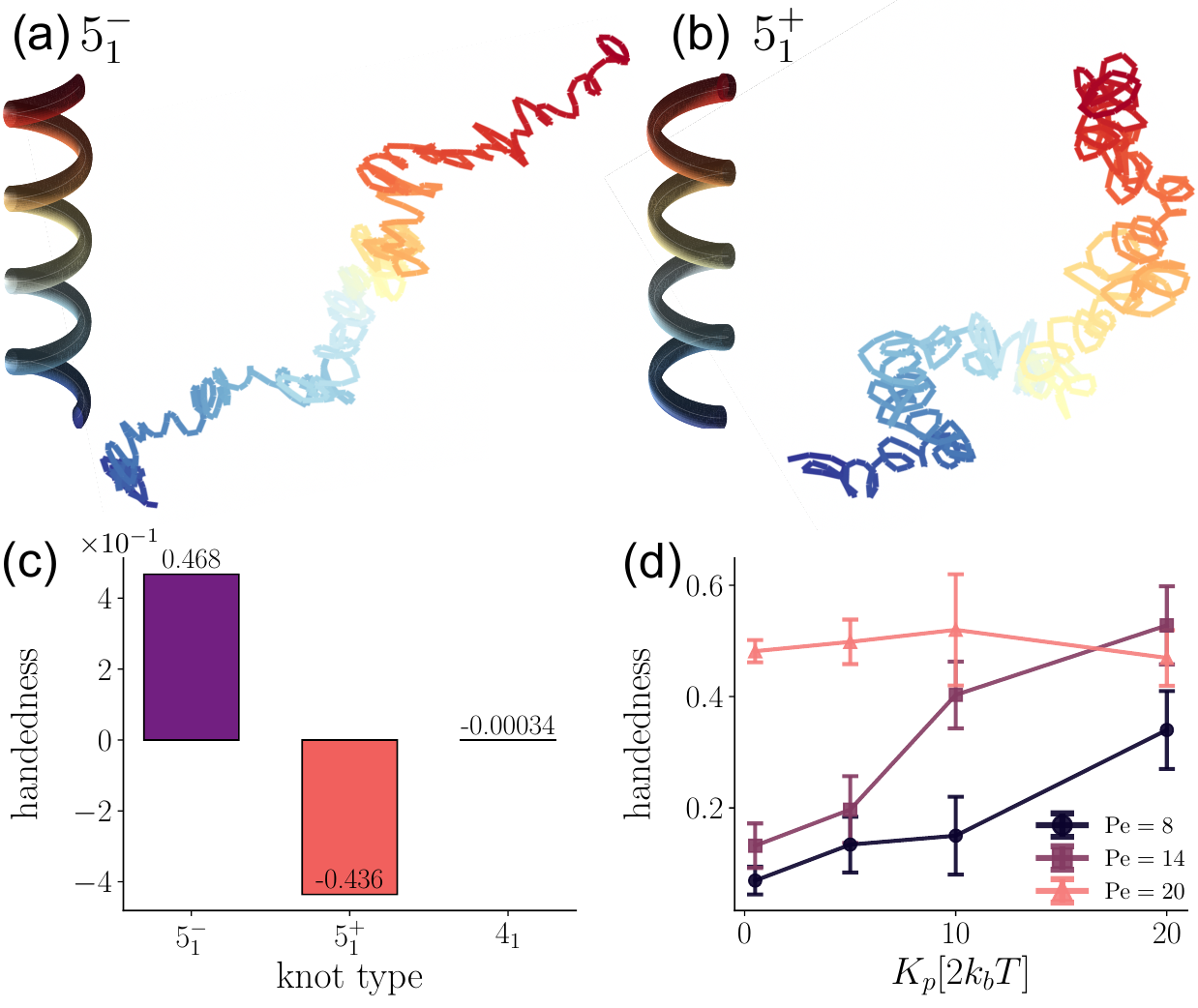}
\caption{\textbf{Chirality.} 
\textbf{(a-b)} Trajectory of a crossing of a $5_1^{-}$ (\textbf{a}) and $5_1^{+}$ (\textbf{b}) knot, over an interval of $100\tau$, simulated at $\text{Pe}=20$. \textcolor{black}{Regular helices are schematic illustrations to guide the eye.}
(\textbf{c}) Average handedness of the trajectories of a crossing of a simulated $5_1^{-}$ (purple), and a $5_1^{+}$ (orange) knots, and of the trajectory of a comoving point to a simulated $4_1$ knot (black bar). (\textbf{d}) Average handedness of comoving trajectories to $5_1^{-}$ knots simulated at $\text{Pe}=14$ with different values of the stiffness parameter $K_p$.
} 
\label{fig3}
\end{figure} 

Another remarkable feature of the dynamics of active knots is illustrated in  Fig.~\ref{fig2}(e-f): the time it takes a bead (highlighted in red for $5_2$ at $\text{Pe}=20$) to traverse the knot is considerably shorter than the timescale 
over which the knot's shape itself changes significantly. Drawing an analogy with polymer reptation in melts~\cite{deGennes1971,Doi1978}, one can picture an active knot spinning within a knotted tube whose shape evolves slowly 
while its topology is preserved. This picture holds for all knots studied.

By tracking either the center of mass of a tight twist knot (panel~\ref{fig2}e) or the crossings of a delocalized torus knot (\textcolor{black}{blue trajectory} in panel~\ref{fig2}f for $5_1$ at $\text{Pe}=20$), we observe that the knot conformation revolves in space. %This reveals two distinct timescales: the reptation period (the time for a bead to complete one lap around the knot) and the knot's revolution period.
\textcolor{black}{This reveals two distinct timescales: the \emph{reptation period}, for a tagged bead to complete one full lap around the ring contour, and the \emph{revolution period}, for a comoving feature of the shape (a crossing) to complete a full turn about the center of mass. Both are measured via a proximity-based analysis of the average pitch of the corresponding trajectories (see SI~\cite{refSI}).}

These timescales are reported in Fig.~\ref{fig2}g for the same case as panel~(f). The revolution period of the crossing trajectory is a factor of $10$ larger than the reptation period of the bead, implying that the bead traverses the knot $10$ times per knot revolution.
%It is worth remarking that the described reptation dynamics could be a concurrent cause to the tightening in twist knots, as it could trigger a mechanism reminding of how the active region of part passive-part active diblock polymers, pulls and tighten a knot in the passive region~\abn{add Enzo reference}

%\abn{If we want, we can mention that the reason the correlation for a tight knot is oscillating is that the shape of the knot is spinning, so the barycenter itself is spinning.}

%\textbf{Left and right handed trajectories of active chiral knots}
\subsection{Dynamic chirality in active knots}
Having explored how topology and activity interplay to create distinct conformational and dynamical regimes, we now turn to another fundamental topological feature of an active knot, namely its chirality.
Most knot types cannot be continuously deformed into their mirror images and thus exist in left-handed $(-)$ and right-handed $(+)$ forms. Exceptions are achiral (or amphichiral) knots, whose mirror images are topologically equivalent. Of the knots considered here, $4_1$ and $6_3$ are achiral, while all others are chiral. Inspecting the trajectories of the knots by chirality reveals that this property reverberates through their dynamics as follows. As described previously, delocalized torus knots and localized twist knots revolve around their center of mass, which maintains a fixed spatial direction for times increasing with activity. Consequently, a point rigidly attached to the knot shape, such as a crossing in a highly active torus knot, traces a spiral trajectory in space, as illustrated in Fig.~\ref{fig3}(a-b). Strikingly, spiral motions driven by $5_1^{-}$ and $5_1^{+}$ knots display definite but opposite chirality (panels 3a and 3b). 
\textcolor{black}{We quantify trajectory chirality through a local handedness order parameter $h(t)$ defined from the geometry of the crossing trajectory (see SI).} 

%We quantify this observation by computing a local handedness $h(t)$ at each trajectory point. Following the procedure detailed in the SI~\cite{refSI}, we associate a triad of vectors describing how the crossing trajectory bends relative to the center-of-mass trajectory. The scalar triple product of this triad yields $h(t) = +1$ for right-handed and $h(t) = -1$ for left-handed configurations.
 %\begin{equation}
 %   h(t)=
 %   \begin{cases}
 %   +1 &\text{if right-handed}\\
 %   -1 &\text{if left-handed}
 %   \end{cases}\;,
%\end{equation}

\begin{figure}[t!]
\centering
\includegraphics[width=1.0\columnwidth]{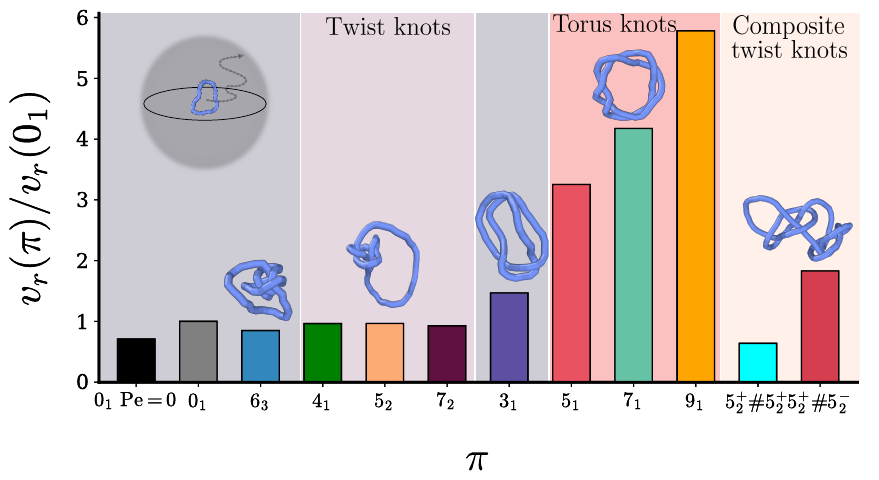}
\caption{\textbf{Topology and efficiency.} 
Average radial velocity for various knot types $\pi$ relative to the unknot ($0_1$) at $\text{Pe}=20$. As shown in the top-left inset, this is calculated from the first passage time out of a spherical boundary of radius $r=25\sigma$. For comparison, the result for the unknots at $\text{Pe}=0$ is reported too (black bar).
Note that the composite knots are twice as long ($N_b=128$) and stiffer ($K_p=10 k_bT$) than the prime knots.
} 
\label{fig4}
\end{figure} 

The average handedness of the trajectory of the crossings~\cite{refSI} of $5_1^{+}$ and $5_1^{-}$ knots is reported in Fig.~\ref{fig3}c for $\text{Pe}=20$ (orange and purple bars, respectively) and indicates that the left-handed $5_1^{-}$ follows a predominantly right-handed trajectory, whereas the right-handed $5_1^{+}$ knots exhibit the opposite trend. This behavior is reminiscent of chirality-dependent sedimentation observed in 3D-printed knots~\cite{Weber2013,Magdalena2018}. Here, however, persistent self-propulsion emerges from the interplay of topology, activity, and chirality.
\begin{figure*}[t!]
\centering
\includegraphics[width=2.05\columnwidth]{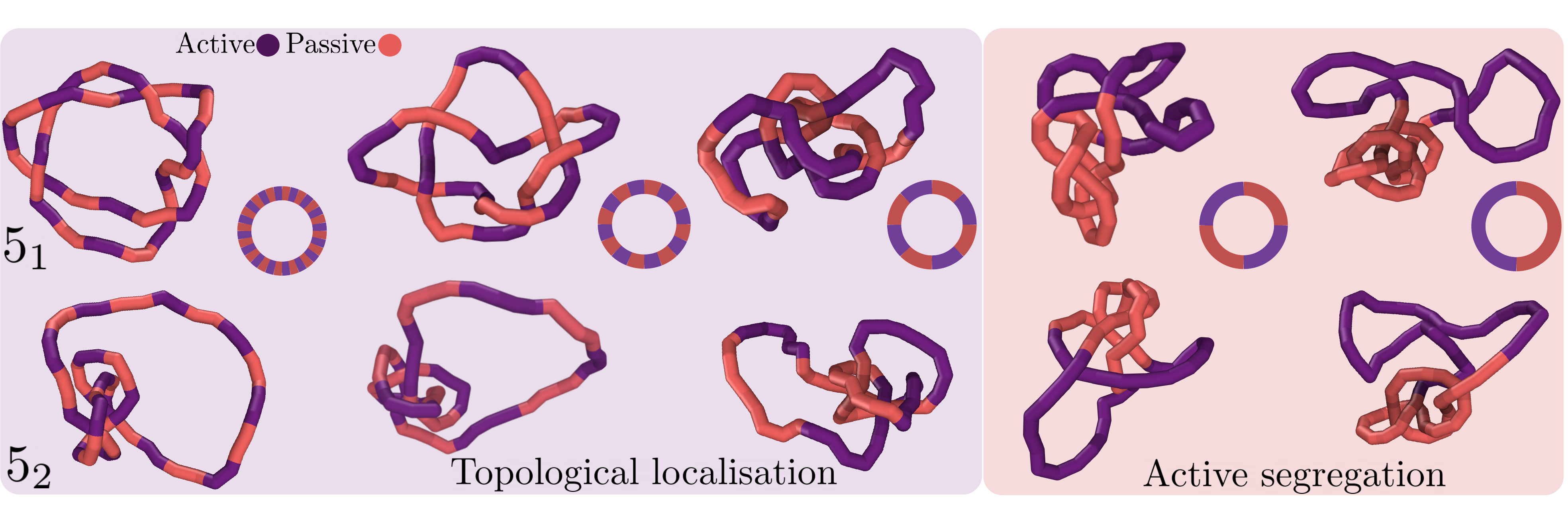}
\caption{\textbf{Activity patterning.} 
Snapshots from simulations of active knots for different knot types $\pi$, $5_1$ (upper row), and $5_2$(bottom), at $\text{Pe}=12$, for different activity patterning (circles in the middle).  
} 
\label{fig5}
\end{figure*} 
We also analyzed achiral $4_1$ knot trajectories at $\text{Pe}=20$. Because $4_1$ conformations deviate significantly from quasi-flat geometries, resolving their crossings is difficult. Decoupling the knot's shape trajectory from high-frequency individual bead rotation thus requires a more general approach. As detailed in the SI, we characterize shape evolution by identifying comoving trajectories, namely time sequences of beads that move alongside the knot image.
The $4_1$ knot fails to maintain persistent chirality, showing near-zero net handedness (Fig.\ref{fig3}(c), black bar).

To understand how baseline rigidity affects handedness and trajectory persistence, we varied the stiffness parameter $K_p$ at different activity values. Results for a $5_1^{-}$ knot at $\text{Pe}=20$, $14$ and $8$ are shown in Fig.\ref{fig3}(d). As $K_p$ increases and the knot inflates toward ideal conformations, trajectory handedness stabilizes. This stabilization is more pronounced at lower activity since activity itself enhances effective knot stiffness.

Finally, to emphasize how topology influences the self-propulsion efficiency of soft rings, we measured the escape rate of various active knots from a spherical boundary, normalized by the unknot's rate. The results in Fig.\ref{fig4} show that torus knots are the most efficient self-propellers. Their escape speed increases nearly linearly with crossing number, with the $9_1$ knot escaping approximately six times faster than the unknot.

 Torus knots develop coherent alignment of active forces, which promotes knot delocalization and persistent self-propulsion, whereas twist knots frustrate this alignment and remain localized. As a consequence, topology controls not only knot morphology but also transport properties and dynamical chirality, with chiral torus knots behaving as persistent self-propelled quasiparticles whose handedness is fixed by topology.

\subsection{Topological localisation vs active segregation}
\textcolor{black}{The above findings are for uniformly active polymers but there exist systems where active agents act on the polymer backbone in a non-uniform manner. An %paradigmatic 
example is that of molecular motors, which bind locally to the backbone, consume chemical energy (typically ATP), and inject finite amounts of force directly on the filament. The result is that the polymer itself becomes a heterogeneous active medium. Specific examples include transcription factors on DNA, SMC complexes on chromatin, and myosin on actin. Except for the recently studied active/passive diblock case~\cite{vatin2025upsurge}, the effect of activity heterogeneity has never been addressed so far. We performed simulations with different activity profiles. Figure~\ref{fig5} shows representative configurations of the $5_1$ (top row) and $5_2$ (bottom row) knots at $\text{Pe}=12$ for several activity patterns (represented as circles). The delocalisation of torus knots and the localisation of twist knots persist for regular patterns of activity with sufficiently spatial small period along the chain. As the length of the active and passive segments increases, however, the knot begins to segregate into the passive region, as observed in diblock copolymers~\cite{vatin2025upsurge}. We term this \textit{active segregation}, to distinguish it from the topological localisation set by the knot type itself, appearing only for twist knots.}

\section{Discussion} \textcolor{black}{In conclusion, we have shown that topology governs the organization of active stresses in knotted rings. 
Torus knots develop coherent alignment of active forces, which promotes knot delocalization and persistent self-propulsion, whereas twist knots frustrate this alignment and remain localized. Hence, topology controls not only knot morphology but also transport properties and dynamical chirality, with chiral torus knots behaving as persistent self-propelled quasiparticles whose handedness is fixed by topology.}

%Torus and twist knots behave asymmetrically: while twist knots localize along the ring, torus knots generically delocalize under activity. This behavior contrasts with passive polymers under tension, where knots typically tighten and localize. We attribute this difference to the distinct internal organization of the two families: torus knots exhibit higher polar alignment of active segments, promoting cooperative propulsion along the backbone and spreading of the knotted region, whereas twist knot geometry disrupts this alignment and stabilizes a compact core.
%\textcolor{red}{We argue that this same alignment mechanism may
%underlie the crumpling transition reported for longer active knots and rings~\cite{breoni2025}, providing a microscopic interpretation of that behavior.}

\textcolor{black}{These results generalize the findings of Ref.\cite{breoni2025}, where distinct responses of torus and twist knots were identified in activity-induced collapse of long knotted chains. The polar-alignment mechanism identified here provides a microscopic interpretation of that behavior and reveals new physics, namely, activity enables also spontaneous self-assembly of a topologically persistent chiral rotor, in which the knotted region acts as a soft quasiparticle driving persistent motion of the entire ring. The resulting trajectories display biomimetic helical motility similar to that observed in active filaments and microorganisms~\cite{carenza2019}, with persistence strongly dependent on topology: torus knots generate highly persistent chiral motion, while unknotted rings and twist knots do not. This is, to our knowledge, the first realization of a topologically chiral active particle. Our results demonstrate that polymer topology can handle the transport properties and dynamical organization of active polymers.}

%Note that we adopt a minimal active-polymer model in which a tangential force of fixed magnitude acts on every monomer. This homogeneous driving describes systems where activity is distributed along the entire contour — such as worms, magnetically driven colloidal rings, or filaments uniformly decorated by motors — as opposed to localised driving by a single machine such as a transcription factor. The complementary regime of localised driving has only recently begun to be explored, for instance through active-passive diblock models~\cite{vatin2025upsurge} in which the active portion can pull on and tighten a knot residing in the passive region.

\textcolor{black}{We further show that the topology-driven localization mechanism is robust for moderate spatial heterogeneity in activity. For finely patterned active/passive profiles, torus knots remain delocalized while twist knots remain localized. However,  when the activity domains become sufficiently large, a competing segregation mechanism emerges, driving the knot into the passive region.}

\textcolor{black}{Our model neglects hydrodynamic interactions. For active polymers, these are known to enhance activity-induced swelling~\cite{martingomez2020}, which would %, if anything, 
quantitatively reinforce the torus-knot delocalisation observed here. Additionally, in the high-activity regime where our central results emerge, conformational properties converge to the free-draining limit, rendering hydrodynamics asymptotically irrelevant~\cite{martingomez2020}. In the localised state, crowding within the dense knotted region should further suppress long-range hydrodynamic interactions, though these may still affect knot diffusivity~\cite{bao2003}. This is consistent with the finding that topology and hydrodynamics act nearly independently in linked ring polymers~\cite{Raucher2020}. Note that hydrodynamic interactions have been shown to delocalise a trefoil knot on a sheared ring~\cite{liebetreu2018}, mirroring the topological delocalisation we find under internal active driving. Whether external flow-mediated and internal self-generated driving share a common delocalisation mechanism remains an interesting open question.}

%\textcolor{brown}{
%A further idealisation of our model is the neglect of hydrodynamic interactions.  For active polymers, hydrodynamic interactions have been shown to enhance rather than suppress the activity-induced swelling~\cite{martingomez2020}, which would if anything reinforce the delocalisation of torus knots that we report; moreover, in the high-activity regime in which our central results emerge, conformational properties converge to those of the free-draining limit, with hydrodynamics becoming asymptotically irrelevant~\cite{martingomez2020}. In the localised state, the dense, tightly knotted region should quench hydrodynamic transport through crowding, rendering long-range interactions progressively less relevant precisely where localisation sets in, while still affecting  the diffusive mobility of the knot~\cite{bao2003}. This expectation is consistent with the finding that topology and hydrodynamic interactions act nearly independently in linked ring polymers~\cite{Raucher2020}. In a related but distinct setting, hydrodynamic interactions have been shown to delocalise a trefoil knot on a sheared ring polymer~\cite{liebetreu2018}, mirroring the topological delocalisation we obtain here under internal active driving. An intriguing open question is precisely how such external driving, mediated by an imposed flow, compares to the internal, self-generated driving of an active knot, and whether the two routes to delocalisation share a common mechanism.  }

\textcolor{black}{An additional important simplification of our model is the representation of activity as point-like tangential forcing. In biological systems, active agents such as polymerases, chromatin remodellers, or molecular motors possess a finite size, and they may interact simultaneously with several nearby polymer segments (for instance, bridging them)~\cite{alberts2015essential}. Finite-size active agents would introduce additional steric constraints within entangled regions and could therefore modify the balance between knot tightening and delocalisation. With respect to experimenatl realisation, our model is therefore best viewed as a coarse-grained description of local active forcing along a filament rather than a detailed representation of any specific molecular motor. The approximation is likely most appropriate for systems with a clear separation between the size of the active unit and the polymer thickness. Biologically, this description may therefore be more appropriate for chromatin subject to ATP-dependent active processes than for naked DNA. It may also be realised in synthetic active filaments functionalised with catalytic or other active inclusions that generate local forcing along the backbone. Investigating how finite-size active agents and more realistic force-generation mechanisms modify the topology-dependent behaviour reported here remains an important direction for future work.
}

\textcolor{black}{Beyond biophysics, our findings suggest a topological design principle for synthetic active matter. Unlike conventional microswimmers, whose motion is programmed through shape or force placement, active knots encode morphology, persistence and trajectory handedness through topology alone. Chiral torus knots therefore, provide a simple realization of soft self-propelled particles~\cite{wei2026microprinting}  whose motility can be selected by knot type.}

%\textcolor{brown}{
%Beyond their biophysical motivation, our results suggest a topological design principle for synthetic active matter and microrobotics. A central goal in that field is to program the motion of microscale units through their geometry and the placement of active forces, an avenue opened in particular by 3D microprinting of anisotropic and deformable active particles~\cite{wei2026microprinting}. Within this effort, shape has been used to encode helical, screw-like propulsion, and toroidal microswimmers have been engineered to transport passive and active cargo~\cite{baker2019microtori}.
%Active knots offer a complementary, purely topological route to the same end: a single flexible ring, driven uniformly along its contour, in which the knot type alone selects whether the object inflates or stays compact, how persistently it moves, and the handedness of its trajectory. Our chiral torus knots, in particular, realise self-propelled  swimmers whose helical handedness is locked to their topology, suggesting that knotting could serve as a programmable and reconfigurable control parameter for soft microrobots, for instance in magnetically actuated colloidal rings or 3D-microprinted filaments.}

\textcolor{black}{An important extension concerns systems containing many interacting knots, either in dense active melts or on the same polymer chain, where topology may generate collective nonequilibrium states and transport—analogous to active topological glasses formed by unknotted rings~\cite{smrek2020}. More broadly, our work identifies active knots as a new class of topological active matter, in which knotted regions behave as self-organized active quasiparticles with tunable morphology and transport properties.}

%Finally, important question concerns the collective behavior of active knot melts, where topology may govern emergent phases and transport—analogous to active topological glasses formed by unknotted rings~\cite{smrek2020}. Another interesting extension is that of eletrophoretic motility of active DNA knots, where it would be interesting to see whether the topology-dependent persistence translates into differential mobilities. Finally, activity may enable interactions between multiple knots on the same chain, potentially creating novel nonequilibrium topological states worth characterizing. More broadly, our results suggest that active knots represent a new class of topological soft matter, where knotted regions act as self-organized active quasiparticles with tunable properties.

\section{Materials and Methods}
We perform molecular dynamics simulations of active bead-spring polymer rings knotted with various fixed topologies. Excluding the case of the composite twist knots $5_2\#5_2$, for which we doubled the length, each simulated knot is made of $N_b=64$ beads of size $\sigma$.
Steric interactions between unbonded beads are modeled by the Weeks-Chandler-Andersen (WCA) potential (Lennard-Jones potential truncated at its minimum) 
\begin{equation}\label{UWCA}
U_{WCA}(r) = \left\{\
\begin{array}{lcl}
4\epsilon \left [ \left ( \frac{\sigma}{r-\Delta}\right )^{12} - \left (\frac{\sigma}{r-\Delta} \right )^6 \right ] + \epsilon & r \leq r_c\\
0 & r > r_c
\end{array} \right. \, ,
\end{equation}
where $r_c=2^{1/6}\sigma+\Delta$, $\epsilon=2k_bT$ and $r$ is the distance between interacting beads.
Consecutive beads are strung through the Finite Extensible Nonlinear Elastic (FENE) potential
\begin{equation*}\label{UFENE}
\begin{split}
U_{\rm FENE}(r) =& U_{WCA}(r)  \\
+&\left\{
\begin{array}{lcl}
-0.5kR_0^2 \ln\left(1-\left(\frac{r-\Delta} { R_0}\right)^2\right) & \ r\le R_0 \\ \infty & \
r> R_0 &
\end{array} \right. \, ,
\end{split}
\end{equation*}
with $k=30\epsilon/\sigma^2$, $R_0=0.4\sigma$, and $\Delta=0.5\sigma$. We chose a small $R_0$ to preserve topology across all cases considered, and verified that the results are insensitive to the specific parameter choices provided that topology is maintained. Bending rigidity is controlled via Kratky--Porod potentials acting on the angle $\theta$ between triplets of consecutive beads,
\begin{equation}\label{UKP}
U_{KP}(\theta) = K_p\left(1+\cos(\theta)\right).
\end{equation}
In most of our simulations we set $K_p=1\epsilon$, corresponding to a persistence length of $2\sigma$, but, by varying $K_p$ we also explore how the baseline stiffness affects the motility of active knots (see main text and below).

The beads of the knots are assigned an orientation, so that bead $i$ is bonded to beads $i-1$ and $i+1$ (with $0$ bonded to $N_b-1$ and $1$, and $N_b-1$ to $N_b-2$ and $0$), and activity is modeled as a constant tangential force, of strength $F_a \in [0,10]\frac{\epsilon}{\sigma}$, acting on every bead:
\begin{equation}
\bm{F}_{\bm{a},i}(\bm{r}_i,\bm{r}_{i+1})=F_a\frac{\bm{r}_{i+1}-\bm{r}_{i}}{||\bm{r}_{i+1}-\bm{r}_{i}||},
\end{equation}
where $\bm{r}_i$ is the position of the $i$-th bead.
The equations of motion are evolved by using LAMMPS in the NVE ensemble coupled to a Langevin thermostat, with integration timestep $dt=10^{-3}\tau$.
Specifically, the beads of the knots evolve according to the following set of Langevin equations:
\begin{equation}\label{LANGEVIN}
m \dfrac{d^2 \bm{r}_i}{dt^2} = -\zeta \dfrac{d \bm{r}_i}{dt} - \nabla U_i + \sqrt{2 k_BT \zeta} \bm{f}_i +\bm{F}_{\bm{a},i},
\end{equation}
where $\zeta$ is the friction, $U_i= U_{WCA}(r_i)+U_{KP}(\theta_i)+U_{\rm FENE}(r)$ and $\bm{f}_i$ is Gaussian white noise. The time unit is $\tau=\sigma \sqrt{m/\epsilon}$, where $m$ is the mass of a bead, and we call $\Delta t=10\tau$ the typical snapshot time interval. 

To prepare the initial configurations, we perform BFACF \cite{Berg1981,Aragao1983Physique,Aragao1983} moves, which are known to preserve knot and link types, on handcrafted knotted polygons on the cubic lattice of spacing $1 \sigma$, until the contour length reaches a desired value.
Following this initial step, we obtain the starting configuration for MD simulations by replacing each lattice vertex of the polygonal knot with a bead of size $\sigma$. We then equilibrate the system for a sufficiently long time before switching on the activity.

%\fi
%\begin{figure}[htbp]
%\centering
%\includegraphics[width=0.6\columnwidth]{handedness_running_ave.pdf}
%\caption{handedness con running average. $h_{ave}(t)=\sum_{i=-10}^{10}h(t+i)/20$}. 
%$h=\begin{cases}+1\;\text{right}\\-1\;\text{left}\end{cases}$ is the handedness.
%\label{fig4tmp}
%\end{figure} 
%\fi

\bibliography{biblio.bib}
\textbf{Acknowledgments}
We thank EPSRC for access to the HPC resources at EPCC (Cirrus).\\
\textbf{Author contributions} A.B., D.M., E.O. and G.N. designed research. A.B. and G.N. performed numerical simulations.  A.B., D.M., E.O. and G.N. wrote the paper.\\
\textbf{Competing interests} The authors declare that they have no competing interests. \\
\textbf{Data and
materials availability} All data needed to evaluate the conclusions in the paper are present in the paper and/or the Supplementary Materials. All simulation code, and data generated by this code, used in this manuscript are available on GitHub \url{https://github.com/GNegroLab/Active-Knots/tree/main} under an MIT license.

\end{document}